\documentclass[11pt,fleqn]{article} 
\usepackage[dvips]{graphicx,psfrag,overpic,color}
\usepackage{latexsym}
\usepackage{amssymb,amsmath,amsfonts,amsxtra,amstext}
\usepackage{rotating,tabularx,multicol,fancyhdr}

\title{Security problems with a SC-CNN-based Chaotic Masking Secure
Communication System \footnote {Submitted to International Journal of Bifurcation and Chaos}}
\author{A. B. Orue, G. Alvarez, F. Montoya\\
\small Instituto de F\'{\i}sica Aplicada, C.S.I.C.\\
\small c/. Serrano 144, 28006 Madrid, Spain\\
\small \texttt{abol@imaff.cfmac.csic.es}\\
\and C. Sanchez-Avila\\
\small Dep. Matem\'{a}tica Aplicada a las Tecnolog\'{\i}as de la Informaci\'{o}n,\\
\small E.T.S.I. Telecomunicaci\'{o}n\\
\small Universidad Polit\'{e}cnica de Madrid, Madrid 28040, Spain
}
\date{}

\begin{document}
\maketitle
\begin{abstract}
This paper studies the security of a chaotic cryptosystem based on the Chua circuit and implemented with State Controlled Cellular Neural Networks. It is shown that the plaintext can be retrieved by ciphertext band-pass filtering after an imperfect decoding with wrong receiver parameters. It is also shown that the key space of the system can be notably reduced easing a brute force attack. The system parameters were determined with high precision through the analysis of the decoding error produced by the mismatch between receiver and transmitter parameters.
\end{abstract}


\bibliographystyle{unsrt}
\sloppy
\parskip=8pt

\section{Introduction}

The possibility of synchronization of two coupled chaotic systems
was first shown by Pecora and Carrol~\cite{Pecora90,Pecora91}, due to the nonpredictable behavior of chaotic variables, it was soon
envisaged the possibility of using them in the field of secure
communications in the same way as the withe noise and random
sequences were used in classical cryptography, accordingly, a great number of cryptosystems based on chaos have been
proposed~\cite{Cuomo93a,Cuomo93b,Wu93,Lozi93,Perez95,Li04,Yang04};
some of them fundamentally flawed by a lack of robustness and
security~\cite{Perez95,Alvarez04,Alvarez05a}.

The Chua circuit~\cite{Chua92,Chua93} is a simple chaotic circuit
considered the paradigm of chaos. It is defined in its dimensionless form by the following state equations:
\begin{align}\label{chua-adimens}
  \dot{x} &=\alpha \left[ y-h(x)\right],\nonumber \\
  \dot{y} &= x-y+z, \\
  \dot{z} &= -\beta y,  \nonumber
\end{align}
with $h(x)=m_1 x + 0.5(m_1-m_0)(|x+1|-|x-1|)$, where $x$, $y$ and
$z$ are the system variables; $\dot x$, $\dot y$ and $\dot z$, are
the derivative the variables with respect to the time $\tau$;
$\alpha,\,\beta,\,m_0$ and $m_1$ are the system parameters.

In~\cite{Arena95} a Generalized Cellular Neural Network (CNN) Cell
Model, was introduced, it was shown that it was possible to
implement the Chua circuit using a State Controlled Cellular Neural
Network (SC-CNN) formed by the suitable interconnection of tree
generalized CNN cells, it is defined by the following state
equations:
\begin{align}\label{chu-CNN}
  \dot x_1 &= -x_1+s_{11}x_1+s_{12}x_2+a_1y_1,\nonumber \\
  \dot x_2 &= -x_2+s_{21}x_1+s_{23}x_3, \\
  \dot x_3 &= -x_3+s_{32}x_2+s_{33}x_3\nonumber,
\end{align}
where $y_1=0.5(|x_1+1|-|x_1-1|)$.

It can be seen that Eqs.~\eqref{chua-adimens} are a particular case
of Eqs.~\eqref{chu-CNN} if:\\
$a_1=\alpha(m_1-m_0)$; $s_{11}=1-\alpha \,m_1$; $s_{12}=\alpha$;
$s_{21}=s_{23}=s_{33}=1$; $s_{32}=\beta$.

Recently it was proposed a new chaotic cryptosystem implemented with
State Controlled Cellular Neural Network
(SC-CNN)~\cite{Kilic04,Gunay05}, it was based on chaotic masking
system with feedback algorithm~\cite{Milanovic96a}. The results of a
PSpice simulation were presented in~\cite{Kilic04} and its hardware
implementation was described in~\cite{Gunay05}. The cryptosystem transmitter was defined as:
\begin{align}\label{chu-dimensT}
  \dot x_1 &= -x_1+s_{11}x_1+s_{12}x_2+a_1y_1,\nonumber \\
  \dot x_2 &= -x_2+m(t)+s_{23}x_3, \\
  \dot x_3 &= -x_3+s_{32}x_2+s_{33}x_3\nonumber,
\end{align}
where $m(t)= x_1(t)+s(t)$ is the ciphertext, $s(t)$ is the plaintext
and $t$ is the variable time. It can be observed the ciphertext
$m(t)$ feedback in the in the second equation of the system.

The cryptosystem receiver was defined as:
\begin{align} \label{chu-dimensR}
  \dot x'_1 &= -x'_1+s_{11}x'_1+s_{12}x'   _2+a_1y'_1,\nonumber \\
  \dot x'_2 &= -x'_2+m(t)+s_{23}x'_3, \\
  \dot x'_3 &= -x'_3+s_{32}x'_2+s_{33}x'_3\nonumber,
\end{align}
where $y'_1=0.5(|x'_1+1|-|x'_1-1|)$.

The recovered plaintext $s'(t)$ at the receiver end was calculated
as: $s'(t)=m(t)-x'_1(t)$.

\begin{figure}[t]
\centering
  \includegraphics[scale=1]{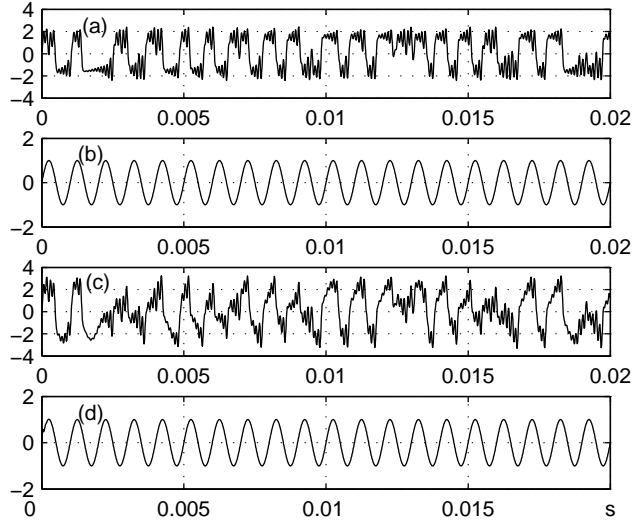} \\
  \caption{(a) Variable $x_1(t)$ of the transmitter; (b) plaintext $s(t)$;
  (c) ciphertext $m(t)=~x_1(t)+~s(t)$;
  (d) retrieved plaintext at the receiver end $s'(t)$.}\label{Fig:criptosignal}
\end{figure}

In~\cite{Kilic04,Gunay05} it was given an example with the following
values: $ \alpha=9$, $\beta=14+\frac{2}{7}, \quad
m_0=-\frac{1}{7},\quad m_1=\frac{2}{7},\quad s(t)=\sin(2\pi 1000\,
t)$, as the circuit was implemented with real resistances and
capacitors of a particular value, it turns out that the time
response of the circuit remain multiplied by a time factor of value
$t/\tau=10^6/51$, with respect to the dimensionless case.

\begin{figure}[t]
\centering
  \includegraphics[scale=1]{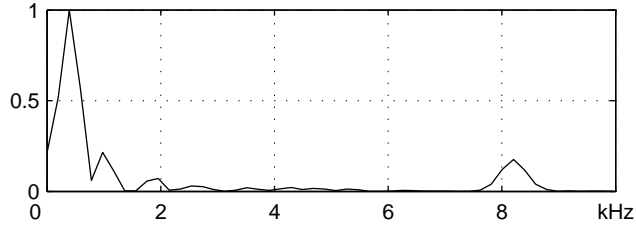} \\
  \caption{Power spectrum of the $x_1(t)$ transmitter variable.}\label{Fig:x1spectrum}
\end{figure}

The signal waveforms of the variable $x_1$, plaintext, ciphertext
and retrieved text are illustrated in Fig.~\ref{Fig:criptosignal}.
In the Fig.~\ref{Fig:x1spectrum} the frequency power spectrum of the
$x_1(t)$ transmitter variable is depicted, it can be seen that most
of the energy is located at the band below 2 kHz, this energy
corresponds high amplitude slow oscillations of $x_1(t)$, there are
also some power components of high frequency, that corresponds to
the small amplitude ripple of $x_1(t)$.

\begin{figure}
\centering
\begin{overpic}[scale=1]{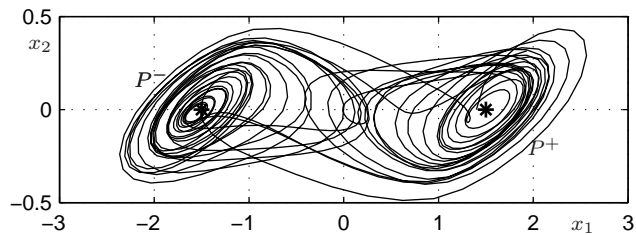}
        \put(83, 12){\scriptsize $P^+$}
        \put(20, 23){\scriptsize $P^-$}
        \put(3, 29) {\scriptsize $x_2$}
        \put(90, 0) {\scriptsize $x_1$}
        \put(20, 23){\scriptsize $P^-$}
\end{overpic}
  \caption{Chua atractor trajectory projected onto the $(x_2,x_1)$ plane.} \label{Fig:x12atractor}
\end{figure}

The Fig.~\ref{Fig:x12atractor} shows the double scroll Chua
attractor formed by the projection on the $(x_2,x_1)$ plane, in the
phase space, of a trajectory portion extending along 0.2 s, for the parameter values of the example in~\cite{Kilic04,Gunay05}. The Chua
attractor trajectory draws two 3D loops, in the vicinity of the
equilibrium points $P^+$ and $P^-$, with a spiral like shape of
steadily growing amplitude, jumping from one of them to the other,
at irregular intervals, in a random like manner. The trajectory may
pass arbitrarily near to the equilibrium points, but never reach
them while in chaotic regime. The two asterisks show the location of
the attractor equilibrium points, of coordinates $x_{1P^\pm}=\pm
(1-\frac{m_0}{m_1})$, $x_{2P^\pm}=0$,
$x_{3P^\pm}=\mp(1-\frac{m_0}{m_1})$; the jumps between loops
corresponds to the low frequency components of the spectrum, while the turns around the equilibrium points corresponds to the high frequency components.

In the present paper it is discussed the weaknesses of this secure
communication system, in the section \ref{sec:definition} the
weakness of the cryptosystem are analyzed and in sections
\ref{sec:filtering} and \ref{sec:parameterdetermina} it is shown how
to break it by filtering and by parameter identification.

\section{Problems with the cryptosystem definition} \label{sec:definition}
Although the authors of~\cite{Kilic04,Gunay05} seemed to base the
security of its cryptosystem on the chaotic behavior of the output
of the Chua circuit, no analysis of security was included and no
indications about key selection, allowable plaintext frequency or
amplitude and system initial conditions.

\subsection{Missing key specification} \label{ssec:key}
The first issue to be considered in a cryptosystem is the secret
key, a cryptosystem cannot exist without a key. When cryptanalyzing
a cryptosystem, the general assumption made is that the cryptanalyst
knows exactly the design and working of the cryptosystem under
study, i.e., he knows every detail about the ciphering algorithm,
except the secret key. This is an evident requirement in today 
secure communications systems, usually referred to as 
principle of Kerchoff~\cite{Stinson95}. In~\cite{Kilic04,Gunay05} it was not
considered whether there should be a key in the proposed system,
what it should consist of, what the available key space would be 
(how many different keys exist in the system), what precision to
use, and how it would be managed. None of these elements should be
neglected when describing a secure communication
system~\cite{Alvarez04b,Alvarez06}.

\subsection{Reduced hypothetical key space}\label{ssec:keyspace}

A typical assumption of most chaotic cryptosystems designers is that the system parameters play the role of key~\cite{Alvarez05a}, such premise will be assumed in the rest of the article.

The simplest strategy for breaking a cryptosystem is known as
\emph{brute force attack} and consist of trying every possible key
on the key space. The attack will be affordable in the case of small key space, hence the number of possible keys must be as huge as possible. Nowadays the veteran Data Encryption Standard is
considered obsolete and abandoned because it has only $7.2\times
10^{16}$ different keys!

The problem of using the Chua circuit as a cryptosystem is that the number of  possible different combinations of useful parameter
values is very short. In~\cite{Matsumoto87,Madan93} it is shown that the Chua circuit exhibit almost every known bifurcation and chaotic phenomenon described in the literature. Its manifold is quite complex, hence it is known as the chaos paradigm. Different
combinations of parameters $\alpha$ and $\beta$ lead to many
different trajectories projected onto the $(x_2,x_1)$ plane, among
them: double-scroll strange attractor, sinks, asymmetric periodic
orbits, period $n$ orbits, R\"{o}ssler like spiral, heteroclinic
orbits, homoclinic orbits and repulsive foci. The only  attractor
behavior suitable for masking the plaintext is the double-scroll
attractor, other behaviors give place to a very simple waveforms that can not hide the plaintext in an efficient manner; but, unfortunately, the region of the $(\alpha, \beta)$ plane giving rise to it is a small fraction of about $4\%$ of all possible combinations of parameter values, as shown in~\cite{Matsumoto87,Madan93}. Hence a hypothetical key space based on the system parameters would be quite small.

This situation is worsened by the fact that some Chua circuit
parameters have a direct relation with the coordinates $x_1$ of the
attractor equilibrium points $P^+$ and $P^-$, that can be
approximately delimited watching at the ciphertext waveform, what
reduces further the key space, as described in the section \ref{sec:parameterdetermina}.

\subsection{Dangerous initial conditions}
For the parameter values of the example given
in~\cite{Kilic04,Gunay05}, there are many unstable periodic orbits
embedded in the double scroll attractor. If for any reason during
the operation of the system some special points are reached, or the
initial conditions include them, the system becomes unstable with
ever growing amplitude of the variables, such points must be
considered forbidden during normal operation, one of this isolated
points is $\{x_1(0),x_2(0),x_3(0)\}=\{1.83487, \;0,\; 2.53784\}$.
A complete forbidden region of the attractor orbit and/or initial
conditions correspond to the values $x_2 \ge 1.08$, for any value of
$x_1$ and $x_3$.

\subsection{Unacceptable plaintexts} \label{ssec:unaceptable}
As the plaintext is feeded with high amplitude into the transmitter
Eqs.~\eqref{chu-dimensT}, the normal behavior of the Chua circuit is
disturbed. It was found that for plaintext frequencies ranging from
4700 Hz to 4970 Hz the attractor orbit remained synchronized with
the plaintext after some miliseconds following the initiation. This
is a very dangerous situation because the ciphertext reveals the
plaintext.

If the plaintext frequency is comprised between 4970 Hz and 12500 Hz
a unstable periodic orbit of about 9500 Hz takes place, hence the
system is not operable for plaintexts with frequencies inside this
margin. It must be concluded that the acceptable plaintext frequency margin must be limited between 0 Hz and less than 4700 Hz, i.e. it seems that it is designed as to encipher speech signals.

\begin{figure}[ht]
\centering
  \includegraphics[scale=1]{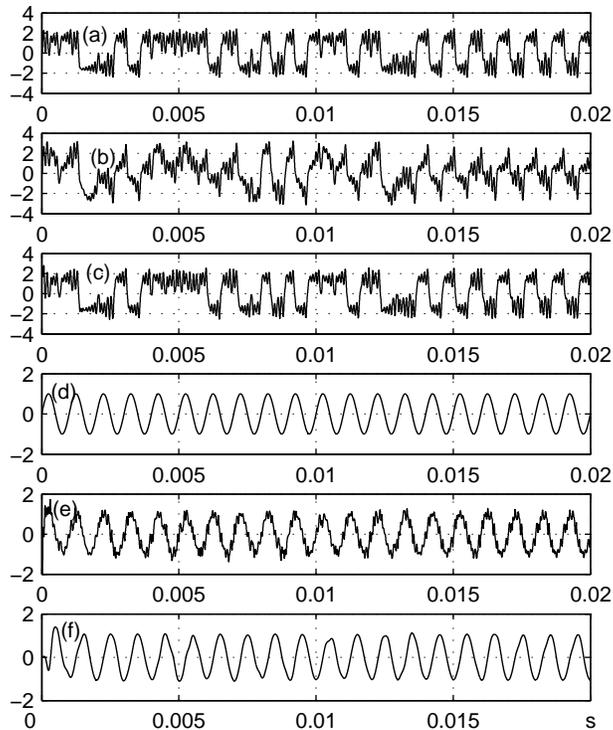}\\
  \caption{Plaintext retrieving with wrong parameter guessing; (a) transmitter chaotic variable $x_1(t)$;
  (b) ciphertext $m(t)=~x_1(t)+ s(t)$;
  (c) receiver chaotic variable $x'_1(t)$;
  (d) plaintext $s(t)$;
  (e) retrieved plaintext $s'(t)$;
  (f) bandpass filtered retrieved plaintext $s'(t)$.}\label{Fig:criptoanalisis}
\end{figure}

\section{Breaking the system by bandpass filtering} \label{sec:filtering}
The main problem with the cryptosystem proposed
in~\cite{Kilic04,Gunay05} is that the synchronism mechanism between
transmitter and receiver is excessively robust, as a result of a
design oriented to avoid the problems presented by the customary
chaotic masking systems. The consequence is that an almost correct
synchronism can be reached for an infinite number of receiver
parameter combinations, for each given transmitter parameter set.

The Fig.~\ref{Fig:criptoanalisis} illustrates this problem. The
system signals are depicted for the transmitter parameter values of
the the example in~\cite{Kilic04,Gunay05} $\alpha=9$,
$\beta=14.2857$, $m_1=0.2857$, $m_0=-0.1428$; and an arbitrary chosen
set of receiver parameters values, far away from those of the
transmitter: $\alpha'=17$,  $\beta'=23.3$,  $m_1'=0.1366$, $m_0'=-m_1'/2$.

It can be appreciated that the waveform of the receiver chaotic
variable $x'_1(t)$ resembles pretty much the corresponding one
$x_1(t)$ of the transmitter. Hence the retrieved plaintext $s'(t)$
differs from the original plaintext $s(t)$ mainly in the high
frequency components, i.e. the jumps between the equilibrium points
are alike, but the rate and amplitude of turns around them are
different, which causes a high frequency noise on the retrieved
plaintext. This noise can be easily removed by filtering. The
Fig.~\ref{Fig:criptoanalisis}~(f) shows the recovered plaintext
after filtering it with a finite impulse response digital bandpass
filter, with 200 taps and a frequency response of 300~Hz to 3.400~Hz, which is the typical bandwidth of telephone loops.

\section{Breaking the system by parameter determination} \label{sec:parameterdetermina}
A possible way to break the system is the brute force attack, which  consists of trying all the possible values of its parameters, until a meaningful and noise-free plaintext is obtained. The number of different combinations could be significantly high, therefore the time needed to try all of them could be enormous, making the envisaged task unattainable. Fortunately the parameters search range may be dramatically reduced in various ways, first by previous study of the cryptosystem characteristics and, second, analyzing the ciphertext waveform.

\subsection{Key space reduction} \label{keyreduction}
\begin{figure}[t]
\centering
\begin{overpic}[scale=1]{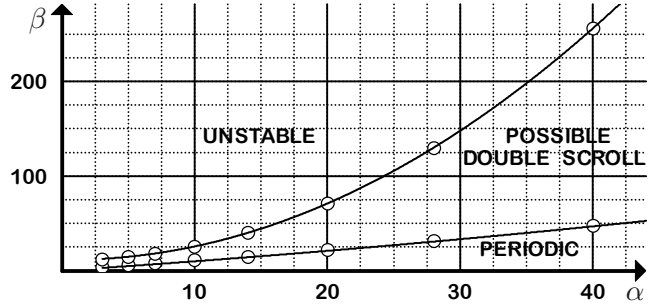}
        \put(3, 43){$\beta$}
        \put(96, .5){$\alpha$}
\end{overpic}
  \caption{Region of the $(\alpha, \beta)$ plane giving rise to the double-scroll attractor.}
  \label{Fig:doblescroll}
\end{figure}

In section~\ref{ssec:keyspace} it was mentioned that the region of the $(\alpha, \beta)$ plane giving rise to the double-scroll attractor in a Chua circuit is a small fraction of all possible combinations of parameter values $\alpha$ and $\beta$. As the system~\cite{Kilic04,Gunay05} differs from the ordinary Chua circuit in that it makes use of feedback, it may have a different behavior than this one; hence the region of the $(\alpha, \beta)$ plane giving rise to the double-scroll attractor was experimentally investigated for different combinations of $m_0$ and $m_1$ values, the results are depicted in the  Fig.~\ref{Fig:doblescroll}, the points inside this region may cause or not a double scroll attractor, depending on the values of $m_0$ and $m_1$, but the points outside this region never cause a double scroll attractor for any combination of $m_0$ and $m_1$ values, therefore they are not suitable for hiding information and need not be investigated when mounting a brute force attack; the region that must be investigated is approximately delimited by the curves $\beta=0,0062\,\alpha^2+0.92\,\alpha+.5$ and $\beta=0,157\,\alpha^2-0.16\,\alpha+12$. Hence the usable key space is notably reduced.

Correspondingly, the region of the $(m_0, m_1)$ plane giving rise to the double-scroll attractor may also be delimited from the Chua circuit definition and from the ciphertext as follows.

According to the definition the Chua system the parameters $m_0$ and $m_1$ are defined as $m_0=(G_a/G)+1$ and $m_1=(G_b/G)+1$, were $G$ is a positive conductance while $G_a$ and $G_b$ are the two negative conductances of the equivalent circuit of the Chua's nonlinear resistor, they satisfy the relation $G_a< G_b<0$, hence it follows that $1>m_1>m_0$. If the  coordinates of the attractor equilibrium points $x_{1P^\pm}=\pm(1-\frac{m_0}{m_1})$ could be determined a tighter relationship between $m_0$ and $m_1$ could be established.

\begin{figure}[t]
\centering
\begin{overpic}[scale=.92]{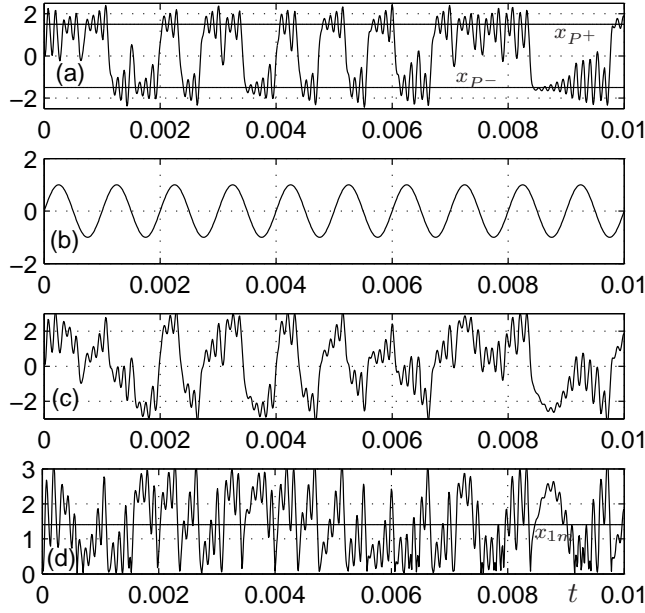}
        \put(83, 87.3){\scriptsize $x_{P^+}$}
        \put(68, 80.5){\scriptsize $x_{P^-}$}
        \put(80, 11)  {\scriptsize $x_{1m}$}
        \put(85, 1.3){$t$}
\end{overpic}
  \caption{Equilibrium points estimation;
  (a) transmitter chaotic variable $x_1(t)$ with $x_{1P^+}$ and $x_{1P^-}$;
  (b) plaintext $s(t)$;
  (c) ciphertext $m(t)=~x_1(t)+ s(t)$;
  (d) absolute value of ciphertext $|m(t)|$.}
  \label{Fig:puntofijo}
\end{figure}

If the undisturbed transmitter chaotic variable  would be accessible, the coordinate $x_{1P^\pm}=\pm (1-\frac{m_0}{m_1})$ of the equilibrium points $P^{\pm}$ could be determined from the variable waveform.  Figure~\ref{Fig:puntofijo}\,(a) shows the waveform of $x_1(t)$ and the true values of $x_{1P^+}$ and $x_{1P^-}$, corresponding to the example given in~\cite{Kilic04,Gunay05}; as can be seen it is not a difficult task to approximate the value of $x_{1P^+}$ or $x_{1P^-}$ as the equidistant line between the relative maxima and minima of the positive part or the negative part, respectively, of the $x_1(t)$ waveform.

But as the only accessible data to an opponent cryptanalyst is the ciphertext $m(t)=~x_1(t)+~s(t)$, depicted in Fig.~\ref{Fig:puntofijo}\,(c), the transmitter variable $x_1(t)$ remains obscured by the presence of the plaintext, so only a coarse estimation of $x_{1P^\pm}$ can be attained, never the less the value may be delimited effectively, establishing two easily measurable bounds. As $x_{1P^+}=-x_{1P^-}$, it is preferable to work with the absolute value of $m(t)$, represented in Fig.~\ref{Fig:puntofijo}\,(d). The value of $|x_{1P^\pm}|$ can be delimited between the bounds $x_{1max}$ and $x_{1mean}$, being the first one the maximum value of $|m(t)|$ and the second one the mean of $|m(t)|$. It is evident from Fig.~\ref{Fig:puntofijo}\,(d) that $|x_{1P^\pm}|<x_{1max}(t)$ and it was found experimentally, for a large assortment of parameter values and plaintexts, that in any case $|x_{1P^\pm}|>|x_{1mean}(t)|$. In the example of~\cite{Kilic04,Gunay05} the true value of $\frac{m_0}{m_1}$ is $\frac{m_0}{m_1}=-0.5$, hence $|x_{1P^\pm}| = 1-\frac{m_0}{m_1}=1.5$, which is in good agreement with the bounds that were experimentally found $x_{1M}=3.00$ and $x_{1m}=1.41$. This fact allows for an important reduction of the search range of all the possible values of $m_0$ and $m_1$. As $x_{1M}=3.00>\pm(1-\frac{m_0}{m_1})>x_{1m}=1.41$ and  $1>m_1>m_0$, it follows that $1>m_1>0$ and $-0,41m_1>m_0>-2m_1$. Again the key space is additionally reduced.

\begin{figure}[t]
\centering
  \includegraphics[scale=1]{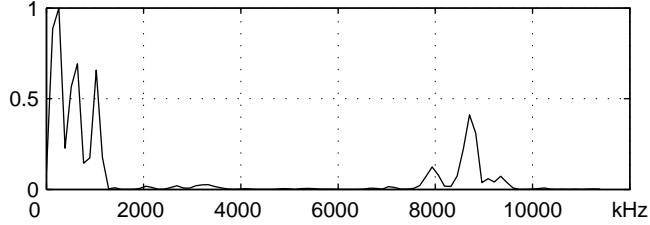}
  \caption{Power spectrum of the receiver decoding error $\varepsilon$, with transmitter parameters: $\alpha=9$, $\beta=14.2857$, $m_0=-0.1428$, $m_1=0.2857$; and receiver parameters: $\alpha'=4.5$,  $\beta'=9$,  $m_0'=-0.12$, $m_1'=0.21$.}
  \label{Fig:noisespectrum}
\end{figure}

\subsection{Parameter determination}
As illustrated in Fig.~\ref{Fig:x1spectrum}, the transmitter variable $x_1(t)$, which acts as a noise to mask the plaintext, has two well differentiated frequency bands. The first one is the low frequency band, with spectral components comprised between $0$ Hz and $3$ kHz, corresponding to the jumps of the atractor between the two loops centered at the equilibrium points $P^+$ and $P^-$, this part effectively conceals the plaintext, that has the same frequency band. The second one is the high frequency band, located beyond $6$ kHz, corresponding to the loops of the attractor trajectory around the equilibrium points.

\begin{figure}[t]
\centering
\begin{overpic}[scale=1]{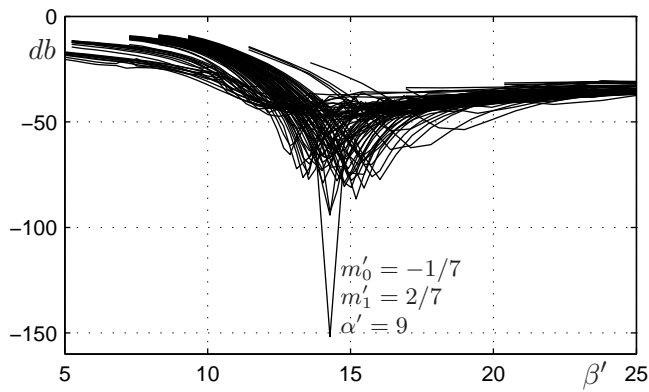}
        \put(90, 0){$\beta'$}
        \put(3, 51){$db$}
        \put(52, 8){\footnotesize $\alpha'=9$}
        \put(52, 12.5){\footnotesize $m'_1=2/7$}
        \put(52, 17){\footnotesize{$m'_0=-1/7$}}
\end{overpic}
  \caption{Power of the high frequency components of the decoding error $\varepsilon$, for different sets of receiver parameters values: $\alpha'=\{4,\ldots , 20\}$;  $m'_1=\{0.01,\ldots , 0.9\}$; $m'_0=\{0.01,\ldots , 1.8\}$}
  \label{Fig:betanoise}
\end{figure}

When the the ciphertext is decoded with an unauthorized receiver, with wrong parameter guessing, the retrieved plaintext  $s'(t)=m(t)-x'_1(t)=s(t)+x_1(t)-x'_1(t)$ is composed by the plaintext and the decoding error $\varepsilon=x_1(t)-x'_1(t)$, that can be considered as an unwanted masking noise. If the receiver and sender parameters were equal the decoding error will disappear, consequently a strategy to retrieve the plaintext may consist of determining the receiver parameters that minimize the decoding error. Unfortunately the noise and the plaintext share the low frequency band of the spectrum, hence it is impossible their complete separation; but still it is possible to separate the high frequency band of the decoding error.

The Fig.~\ref{Fig:noisespectrum} illustrates the power spectrum of this error, the low frequency components are mixed with the plaintext; but the high frequency components are far from the plaintext frequencies, therefore the decoding error created by the high frequencies of $\varepsilon$ can be easily extracted from the ciphertext by means of high-pass filter with cut-off frequency of $6.5$ kHz.

\begin{figure}[t]
\centering
\begin{overpic}[scale=1]{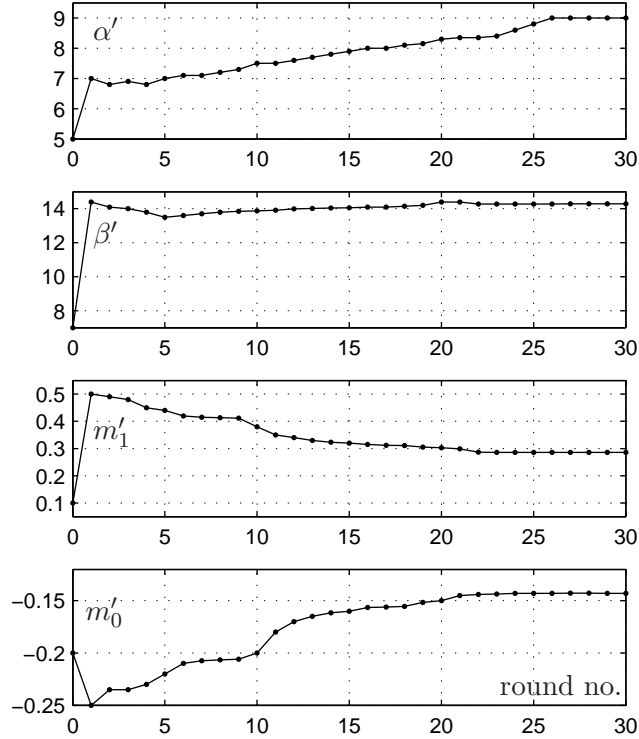}
        \put(11, 93){$\alpha'$}
        \put(11, 66){$\beta'$}
        \put(11, 40){$m'_1$}
        \put(10, 16){$m'_0$}
        \put(65, 6){round no.}
\end{overpic}
  \caption{Story of the parameters value approximation.}
  \label{Fig:aprox}
\end{figure}

The Fig.~\ref{Fig:aprox} illustrates the logarithm of the power of the high frequency components of the decoding error $\varepsilon$, for different sets of receiver parameters $\alpha'$, $m'_1$ and $m'_0$ in function of the $\beta'$ parameter. It can be seen that the minimum decoding error is reached when the transmitter and receiver parameters agree. All the curves show the same tendency: the decoding error grows with the mismatch between the transmitter and receiver parameters, their relative minima is reached for values near to the values of the corresponding parameters in the transmitter.

To determine the parameter values an iterative optimization procedure was developed; it consisted of a number of approximation rounds, in each round the four parameters were varied one each time, looking for the minimum decoding error; the first round started from a set of arbitrary parameters $\alpha'=5$, $\beta'=7$, $m'_1=0.1$, $m'_0=0.2$, it were tried 31 values of each parameter inside the limited range defined in the section~\ref{keyreduction} and the value giving rise to the lowest decoding error was retained; in each next round the margin of variation of each parameter was progressively reduced; the procedure finished when a stable repeat value of the parameters was reached, the number of required rounds were 30 and the elapsed computing time was 965 seconds, in a $4$ GHz Pentium Dual. The Fig.~\ref{Fig:aprox} illustrates the story of the procedure, sowing the variation of each parameter in function of the round number. The parameters values were determined with a precision of five to six significative digits, allowing for the exact retrieving of the plaintext.

\subsection{Conclusion}
The secure communication system described in~\cite{Kilic04,Gunay05} was studied. It was found that the synchronism mechanism is excessively robust, the consequence is that an almost correct
synchronism can be reached for an infinite number of receiver
parameter combinations, therefore the plaintext can be retrieved by simple band-pass filtering after decoding the ciphertext with a receiver with wrong parameters.

It was also found that the key space of the system can be notably reduced by means of the study of the geometric properties and the the chaotic regions of the Chua atractor, making feasible a brute force attack. Finally the parameters of the system were determined with high precision analyzing and minimizing the decoding error created by the mismatch between receiver and transmitter parameters.

\subsection*{Acknowledgment}
This work was supported by \textit{Ministerio de Ciencia y
Tecnolog\`{i}a} of Spain, research grant SEG 2004-02418.


\end{document}